\documentclass[runningheads]{llncs}
\usepackage[newfloat,frozencache]{minted}
\usepackage[subtle]{savetrees}
\usepackage{graphicx}
\usepackage{booktabs}
\usepackage{adjustbox}
\usepackage{hyperref}
\usepackage{amsmath}
\usepackage{amssymb}
\usepackage{mathtools}
\usepackage{wrapfig}
\usepackage{lipsum}
\usepackage{amsfonts}
\usepackage{pifont}
\usepackage{fdsymbol}
\newcommand{\cmark}{\ding{51}}%
\newcommand{\xmark}{\ding{55}}%

\def \0bf{{\mathbf 0}}

\title{Carbon Footprint of Selecting and Training Deep Learning Models for Medical Image Analysis}

\author{ Raghavendra Selvan \inst{1,2} \and Nikhil Bhagwat\inst{3} \and Lasse F. Wolff Anthony\inst{1} \and {Benjamin Kanding}\inst{1} \and {Erik B. Dam}\inst{1}}
\authorrunning{R Selvan et al.}
\titlerunning{Carbon Footprint of DL in Medical Imaging}
\institute{Department of Computer Science, University of Copenhagen, Denmark \and
Department of Neuroscience, University of Copenhagen, Denmark 
\and
McGill University, Canada \\
\email{raghav@di.ku.dk}
}

\begin{document}

\maketitle

\begin{abstract} 
The increasing energy consumption and carbon footprint of deep learning (DL) due to growing compute requirements has become a cause of concern. In this work, we focus on the carbon footprint of developing DL models for medical image analysis (MIA), where volumetric images of high spatial resolution are handled. In this study, we present and compare  the features of four tools from literature to quantify the carbon footprint of DL. Using one of these tools we estimate the carbon footprint of medical image segmentation pipelines. We choose nnU-net as the proxy for a medical image segmentation pipeline and experiment on three common datasets. With our work we hope to inform on the increasing energy costs incurred by MIA. We discuss simple strategies to cut-down the environmental impact that can make model selection and training processes more efficient.

\vspace{-0.25cm}
\end{abstract}

\section{Introduction}

{\em ``Global warming of 1.5°C and 2°C will be exceeded during the 21st century unless \underline{deep reductions}\footnote{Underlining by us for emphasis.} in carbon dioxide (CO2 ) and other greenhouse gas emissions occur in the coming decades"}, reads one of the key points in the most recent assessment report of the Intergovernmental Panel on Climate Change (IPCC-2021)\cite{2021ipcc}. Without sounding as alarmists, the purpose of this work is to shine light on the increasing energy- and carbon- costs of developing advanced machine learning (ML) models within the, relatively small but rapidly growing, domain of medical image analysis (MIA). 

The massive progress in computer vision tasks, as witnessed in MIA applications, in the last decade has been enabled by the class of ML methods that are now  under the umbrella of Deep Learning (DL)\cite{lecun2015deep,schmidhuber2015deep}. The progress of DL methods have been accelerated by the access to  big data sources and big compute resources. According to some estimates, the compute required for DL methods has been doubling every 3.4-6 months since 2010\cite{amodei2018ai,sevilla2022compute}. However, this increasing compute also results in a proportional increase in demand for energy production. In 2010, energy production was responsible for about 35\% of the global anthropogenic green house gas emissions\cite{change2014mitigation}. The broad adoption and success of DL, while exciting, can also evolve into becoming a significant contributor to climate change due to its growing energy consumption.
\begin{figure}[t]
    \centering
    \includegraphics[width=0.5\textwidth]{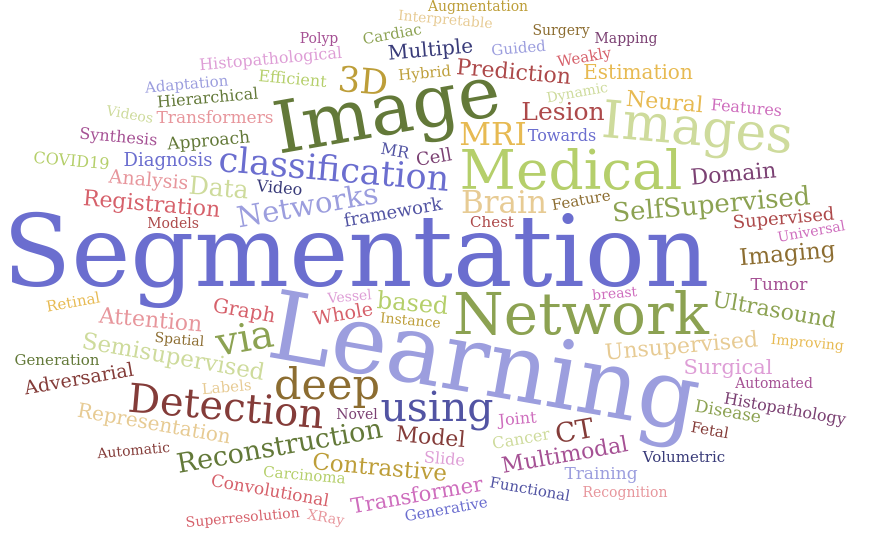}
    \caption{Word cloud showing the trends of papers in MICCAI'21. {\em Deep Learning} and {\em image segmentation} continue to be the dominant trend.}
    \label{fig:cloud}
\end{figure}

In MIA, the ingested data are not usually {\em big} in "sample size" compared to other natural image datasets (e.g. Imagenet\cite{ImageNet}) but they are {\em big data}, in terms of their feature set as they can be volumetric and/or of higher spatial resolution (e.g. 256x256x256 T1-weighted magnetic resonance imaging (MRI) scans). This necessitates novel model customization and selection strategies for model training. Nevertheless, several successes (e.g. U-net\cite{Ronneberger2015}), have generated huge interest in application of DL models in MIA. Particularly, the overall trend of increasing DL for tasks such as segmentation is also captured in venues such as MICCAI\cite{miccai21},
where {\em deep learning} and {\em image segmentation} continue to be the most prominent trends, as captured in the keyword cloud in Figure~\ref{fig:cloud}\footnote{\url{https://www.jasondavies.com/wordcloud/}}. 

In this work, we focus on studying the carbon footprint of DL for MIA using image segmentation as a case study. Segmentation could be a useful representative study, as in MICCAI-2021 there were a total of 143 papers out of the 531 accepted ones with {\em segment} in their title i.e., around 27\% of all papers tackle medical image segmentation in some form or another\cite{miccai21}. We investigate the energy costs and carbon footprint of performing model selection and training the nnU-net segmentation model\cite{isensee2021nnu} on three popular datasets. In line with the MICCAI-2022 recommendations on reporting energy costs of training DL models, we present several methods from literature and compare their features. Finally, we present five simple steps to ensure good practices when developing and training DL models for MIA. 

We present our work as a meta-analytic study to 1) benchmark carbon footprint of DL methods for MIA  and 2) present tools and practices to reduce the compute costs of DL models during future pipeline development in MIA. 






\section{Related Work}
In natural language processing (NLP), where some of the largest DL models are in use, the research community has taken notice of the growing energy consumption and its adverse impact on environmental sustainability and equitable research\cite{strubell2019energy,bender2021dangers}. To draw attention to these issues, initiatives such as dedicated conference tracks focused on sustainable NLP methods are gaining traction\footnote{\url{https://2021.eacl.org/news/green-and-sustainable-nlp}}. There is also growing concern of the environmental impact within bioinformatics where massive data processing is carried out, and their sustainability is being discussed in recent publications focusing on the carbon costs of storage in data banks\cite{samuel2022sustainable} and high performance computing\cite{lannelongue2021carbon,grealey2021carbon}. Closer to the MIA community is the initiative by The Organization for Human Brain Mapping (OHBM) which has instituted a dedicated special interest group (SIG) to focus on the environmental sustainability of brain imaging. Members of this SIG have published an action plan for neuroscientists with suggestions ranging from optimizing computational pipelines to researchers flying less for academic purposes, including to conferences \cite{rae2021climate}\footnote{\url{https://ohbm-environment.org/}}. 

\section{Methods for Carbon Footprinting}
\label{sec:methods}

Any DL model development undergoes three phases until deployment: Model Selection, Training, Inference. Widely reported performance measures of DL models such as the computation time primarily focus on the inference time, and in some cases on the training time of the final chosen model. While these two measures are informative they do not paint the complete picture. Any DL practitioner would agree that the primary resource intensive process in DL is during the model selection phase. This is due to the massive hyperparameter space for searching network architectures, and the resulting iterative model selection process. Moreover, the model selection and training procedures are highly iterative in practice, and many large projects implement them as continual learning. This further underscores the need for better estimation of the recurring costs associated with these procedures. 

The accounting of these costs is contingent on several factors. Reporting the computation time of model development, training and inference can improve transparency in DL to some extent. However, as the computation time is dependent on the infrastructure at the disposal of the researchers, only reporting the computation time provides a skewed measure of performance even when the hardware resources are described. 

On the other hand, reporting the energy consumption (instead of computation time) of the model development provides a more holistic view that is largely agnostic of the specific hardware used. 

\begin{table}[h]
\caption{Feature comparison of four popular carbon tracking tools for ML. The table shows if the tools can be used for tracking (Track.), predicting (Pred.) and visualizing (Vis.) the carbon emissions. Further, if there are options to obtain some form of carbon impact statements (Report), if the tool can be installed as a python package (Pip) and the ease of integrating APIs for extending functionalities are also shown.}
\label{tab:trackers}
\centering
\scriptsize
\begin{tabular}{@{}lccccccc@{}}
\toprule
{\bf Tool} & {\bf Track.} & {\bf Pred.} & {\bf Report} & {\bf Pip} & {\bf Vis.} & {\bf API} & {\bf CPU}
\\ \midrule
MLEC\cite{Lacoste}     &   \xmark     &  \xmark     & \cmark    &  \xmark   & \xmark & \xmark & \xmark 
\\
EIT\cite{Henderson2020} &   \cmark     & \xmark      &  \cmark    & \cmark    & \xmark & \xmark & \cmark 
\\
CarbonTracker\cite{anthony2020carbontracker}     &  \cmark      & \cmark      &   \cmark  &  \cmark& \xmark & \xmark & \cmark
\\
CodeCarbon\cite{codecarbon}     &    \cmark    & \xmark      &   \cmark   &  \cmark   &\cmark & \cmark & \cmark
\\ \bottomrule
\end{tabular}
\end{table}

In the past couple of years, several tools have been developed to help researchers estimate the energy costs and carbon emissions of their model development; four of these with different features are summarized in Table~\ref{tab:trackers} and described below:
\begin{enumerate}
    \item {\bf Machine Learning Emissions Calculator (MLEC)\cite{Lacoste}}: This is a self-reporting interface where users input training time, hardware and geographic location post-hoc to estimate the carbon emissions. It currently only estimates based on the energy consumption of only the graphics processing unit (GPU). 
    \item {\bf Experiment-Impact-Tracker (EIT)\cite{Henderson2020}}: This tool can be used to track the energy costs of CPU and GPU during the model training by embedding few lines of code into the training scripts. The tool can be tweaked to fetch real-time carbon intensity; currently this feature is supported for California (US).
    \item {\bf Carbontracker\cite{anthony2020carbontracker}}: This tool is similar to EIT in several aspects. In addition to the tracking capabilities of EIT, Carbontracker can also predict the energy consumption based on even a single run of the model configuration. This can be useful for performing model selection based on carbon emissions without having to train all the way. This tool currently provides real-time carbon emissions for Denmark and the United Kingdom, and can be extended to other locations. 
    \item {\bf CodeCarbon\cite{codecarbon}}: This is the most recent addition to  the carbon tracking tools and is the best maintained. It provides tracking capabilities like EIT and Carbontracker, along with comprehensive visualizations. The application programming interface (API) integration capabilities are best supported in CodeCarbon which can be useful for extending additional functionalities.
\end{enumerate}
All the four tools are open-source (MIT License), provide easy report generation that summarize the energy consumption and present interpretable forms of carbon emission statements f.x, as distance travelled by car to equal the carbon emissions.\footnote{When the trackers cannot fetch real-time carbon intensity of energy for the specific geographic location, most resort to using some average estimate from a look-up table.}

 \begin{figure}[t]
    \centering
    \includegraphics[width=0.5\textwidth]{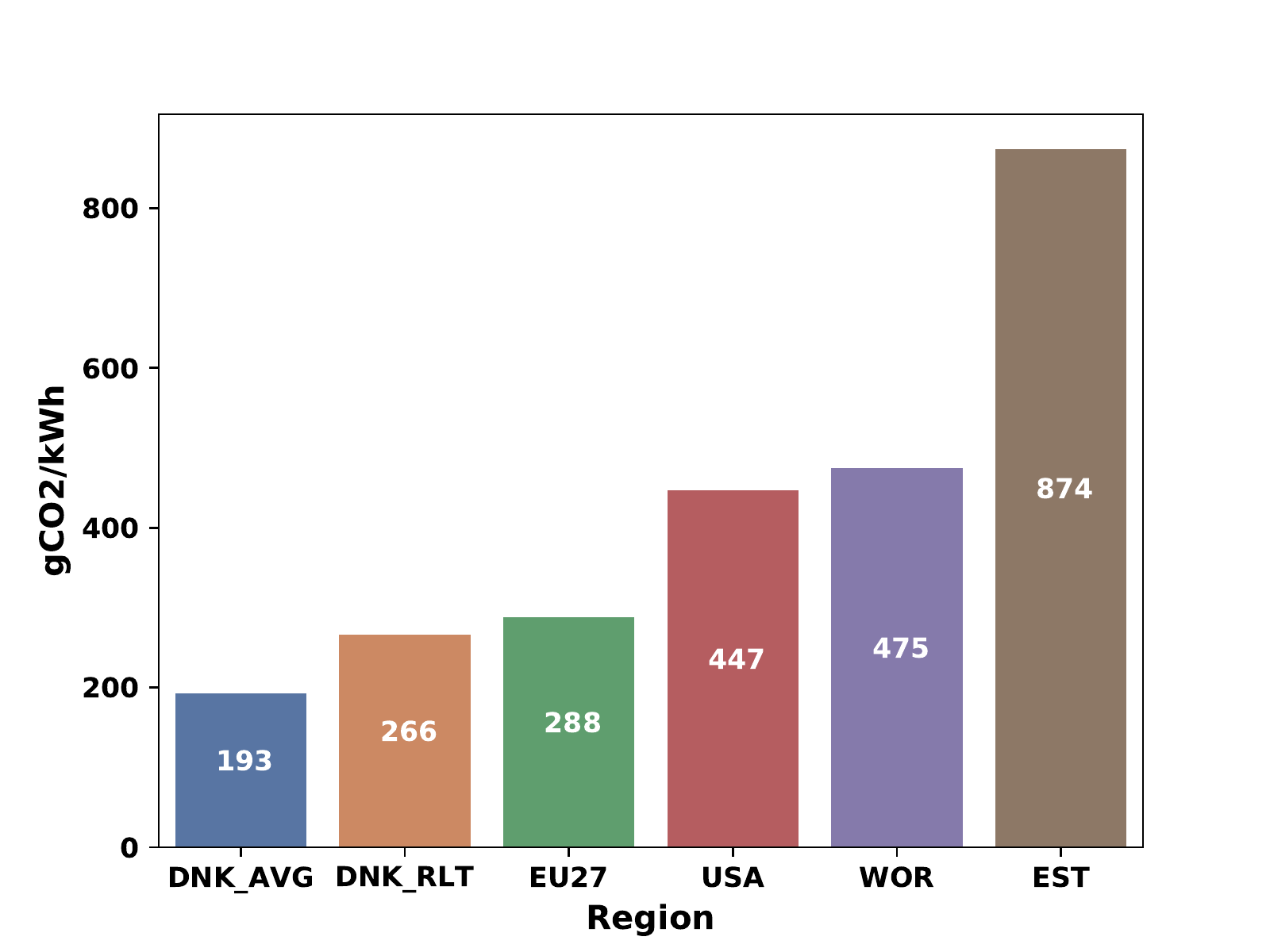}
    \caption{Average carbon intensity over an year, which is the ratio of CO2 emissions from public electricity production and gross electricity production,  for different regions of the world\cite{eea2022co2}. DNK\_RLT is based on the real-time monitoring of the carbon intensity in Denmark over a period of one week in 02/2022. The remaining values are based on the average intensities for Denmark (DNK\_AVG), Estonia (EST) and EU countries (EU-27) as reported by government agencies in the European Union (EU)\cite{eea2022co2} and USA\cite{US2022co2}. The global average (WOR) is obtained from\cite{energy2019co2}. In general, higher ratio of coal powered energy production increases the average carbon intensity, as is evident when comparing DNK\_AVG to EST.}
    \label{fig:avgint}
    \vspace{-0.5cm}
\end{figure}

\section{Data \& Experiments}
\begin{figure}[t]
    \centering
    \includegraphics[width=1.0\textwidth]{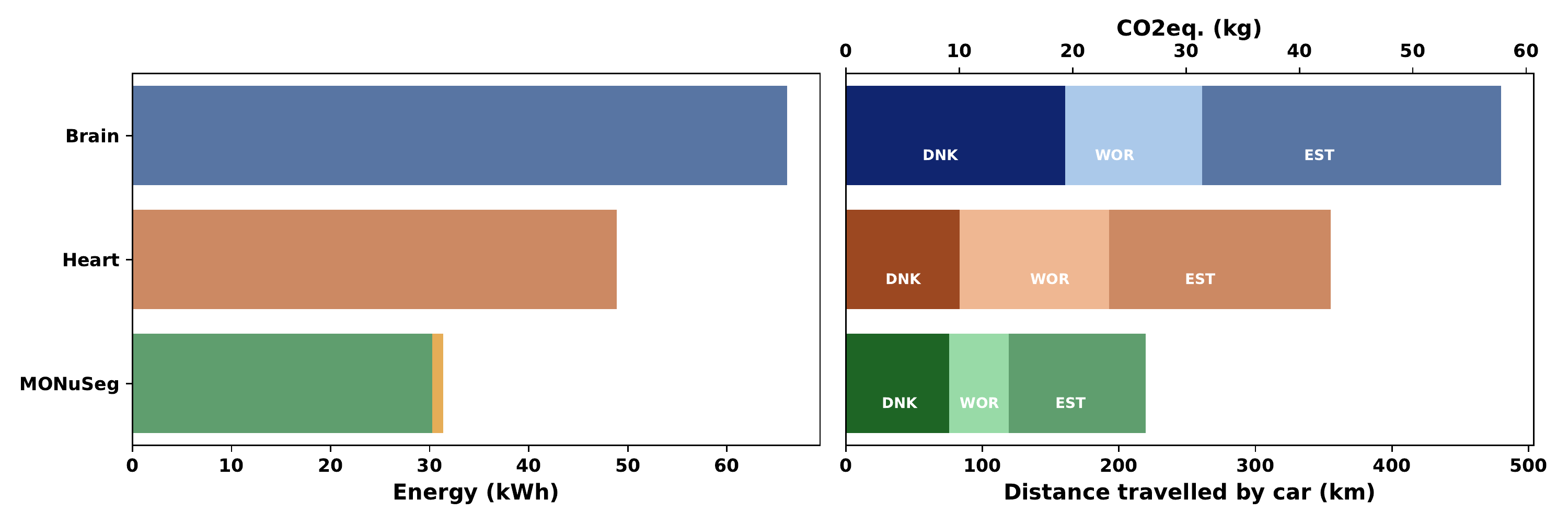}
    \vspace{-0.75cm}
    \caption{(left) Total predicted energy consumed (kWh) over the five-fold cross validation for the three datasets using nnU-net\cite{isensee2021nnu}. For MONuSeg the predicted (orange) and the actual energy consumptions are shown, which are almost the same. (right) Carbon cost due to the training on the three datasets reported in CO2eq.(kg) and equivalent distance travelled by car (km). The carbon intensity and distance are also reported for three geographic  regions (Denmark:DNK, Estonia: EST, Global: WOR) based on the regional average carbon intensities. All measurements were tracked/predicted using Carbontracker\cite{anthony2020carbontracker}.}
    \label{fig:intensity}
\end{figure}

We use medical image segmentation as a case study to present some observations on the energy costs and carbon emissions of performing model selection and training them. We use the recent nnU-net\cite{isensee2021nnu} as the segmentation method as it integrates a comprehensive and automatic framework for segmentation model development, including preprocessing and post-processing steps. \\
{\bf Data:} We present experiments on three diverse MIA datasets for segmentation: 
\begin{enumerate}
    \item {\bf MONuSeg}\cite{kumar2019multi}, which is a multi-organ nuclei segmentation dataset in 2D. The dataset consists of 30 images for training/validation and 14 for testing purposes. Each image is of 1000x1000px resolution and comprising several thousand nuclei annotations per image.
    \item {\bf Heart} segmentation dataset from the medical  decathlon\cite{simpson2019large} and consists of 30 mono-modal MRI scans of the entire heart, with the segmentation target of the left atrium. The dataset is split into 20 for training and 10 for testing.
    \item {\bf Brain} segmentation dataset, which is also from the medical decathlon\cite{simpson2019large} consisting of 4D volumes (multiple-modalities per image) with different types of tumours as the target for segmentation. The dataset consists of 484 volumes for training and 266 for testing purposes.
    \end{enumerate}
{\bf Experimental set-up:}
We use the official version of nnU-net implemented in Pytorch\cite{paszke2019pytorch} and integrate Carbontracker\cite{anthony2020carbontracker} into the network training script to track/predict the energy consumption and carbon emissions, as shown in Appendix~\ref{sec:carbontracker}. All experiments were performed on a desktop workstation with GeForce RTX 3090 GPU with 24GB memory, Intel-i7 processor and 32GB memory. The default configuration of nnU-net uses five-fold cross validation on the training set and a maximum of 1000 epochs per training fold. We follow the K-fold cross validation set-up, with $K=5$ for all datasets and train MONuSeg for 1000 epochs, Heart dataset for 100 epochs and Brain dataset for 50 epochs to reduce the computational cost. We use the predictive model in Carbontracker to report the energy consumption for the full 1000 epochs based on shorter runs\cite{anthony2020carbontracker}. Carbontracker provides real-time carbon emissions for only few regions. To demonstrate the difference in carbon emissions if the experiments were carried out in different geographical regions, we use the average carbon intensity (gCO2/kWh) queried from several sources and are shown along with the sources in Figure~\ref{fig:avgint}. \\
{\bf Results:} The total predicted energy cost of running nnU-net on the three datasets over the five-fold cross validation is shown in Fig.~\ref{fig:intensity} (left). For MONuSeg we have also plotted the predicted energy consumption after the first epoch of training (with transparency and green error bars) and we see there is no substantial difference between the actual- and predicted- energy consumption as estimated from Carbontracker. As the Brain dataset is 4D and has many more data points, the aggregate energy cost is highest compared to the Heart- and MONuSeg datasets.

\begin{table}[t]
\caption{Quantitative comparison of the {\em total} (over all training folds) computation time, energy consumed (in kWh) and carbon costs of different settings reported as the distance travelled by car, for different datasets and different regions. To demonstrate the influence of training precision we compare the mixed and full precision mode training of nnU-net. All experiments are reported for 1000 epochs. For Heart and Brain datasets, they were based on predicted estimates using 100 and 50 epoch runs, respectively. }
\label{tab:distance}
\centering
\scriptsize
\begin{tabular}{@{}l|c|cc|cc|ccc@{}}
\toprule
\multicolumn{1}{c|}{\textbf{Dataset}} & \multicolumn{1}{c|}{\textbf{Comp.Time}}               & \multicolumn{2}{c|}{\textbf{Energy (kWh)}} & \multicolumn{2}{c|}{\textbf{DNK-RLT (km)}}               & \multicolumn{3}{c}{\textbf{Region (km)}}                                                                     \\ 
                                     & & \multicolumn{1}{c}{{\em mixed}} & \multicolumn{1}{c|}{\em full} &  \multicolumn{1}{c}{{\em mixed}} & \multicolumn{1}{c|}{\em full} &
                                     \multicolumn{1}{c}{DNK\_AVG} & \multicolumn{1}{c}{EST} &  \multicolumn{1}{c}{WOR} \\ \midrule
\textbf{MONuSeg}               &  61h 55m    & 30.2 & 35.1&     75.5                 &  80.2                         &         66.8                &         219.6                  &      119.3                                            \\
\textbf{Heart}                  &   116h 40m  &48.9 & 65.8 &       83.2 &  128.5                  &         108.1                  &         355.1                &      192.9                                                                       \\
\textbf{Brain}                   & 368h 20m  &66.1& 84.6&         160.8                 &         170.4                  &      146.1                   &       479.8                    &               260.4                                   \\ \bottomrule
\end{tabular}
\vspace{-0.5cm}
\end{table}

In Fig.~\ref{fig:intensity} (right) and Table~\ref{tab:distance}, the carbon emissions for performing these experiments in different regions (Denmark: DNK, Global average: WOR, Estonia: EST) are reported, along with the corresponding distance travelled by car as reported from Carbontracker. Due to the high average carbon intensity in EST, the carbon emissions are easily three times higher than in DNK or in other words, the carbon emissions in EST are equivalent to driving about 96km compared to 30km for a single nnU-net run on the Brain dataset. 

The default configuration of nnU-net uses automatic mixed precision (AMP) scheme which is known to be efficient by reducing computational overhead by adaptively reducing the quantization (float32 to float16) without significant loss in performance\cite{micikevicius2018mixed}. To investigate the reduction in energy costs due to AMP, we re-run all the experiments for 10 epochs using full precision. We then use Carbontracker to predict the energy consumption and carbon emissions for 1000 epochs. These results are reported in Table~\ref{tab:distance}  where a clear reduction in total energy consumed and the carbon emissions (reported as distance travelled) for all three datasets is noticed, with larger gains in the 3D experiments.

\vspace{-0.25cm}
\section{Discussions}
\vspace{-0.25cm}
In Table~\ref{tab:distance}, we reported the carbon emissions (as distance travelled by car) twice for Denmark: DNK\_RLT and DNK\_AVG, differing based on their average carbon intensity. DNK\_RLT is based on real-time carbon emissions which during the course of the experiments was about 266gCo2/kWh, whereas the DNK\_AVG retrieved from \cite{eea2022co2} for 2018 is 193gCO2/kWh. The difference could be due to the increased loads during winters in Denmark, caused by additional coal powered energy. This highlights the need for {\em tracking} the carbon emissions in real-time instead of self-reporting and also to integrate APIs for live carbon intensity fetching for all the regions into the carbon tracking tools presented in Sec.~\ref{sec:methods}.

We used nnU-net as a proxy for extensive hyperparameter tuning, as it also integrates hyperparameter selection for medical image segmentation. Whenever possible, DL practitioners should use improved {\em hyperparameter optimization} strategies other than grid search.  For instance, simple strategies like random search\cite{Bergstra2012} to more complex methods such as hypernetworks\cite{brock2018smash,hoopes2021hypermorph} can significantly reduce the computation time needed during model selection.

Using {\em energy-efficient hardware} and settings can also help reduce carbon emissions. For instance, power management techniques like dynamic voltage and frequency scaling (DVFS) have been shown to conserve up to ~23.1\% energy consumption by operating at the optimal core frequency instead of the default one\cite{Yang2017}. In instances where the user does not have full control over the hardware configuration, accessing cloud-based hardware can be more efficient as some of the datacenters might have better power usage effectiveness (PUE) than locally maintained hardware.

Figure~\ref{fig:intensity} (right) points to an interesting observation.  Depending on the region where the training is performed, the carbon emissions can vary drastically. This is due to the regional carbon intensity as shown in Figure~\ref{fig:avgint} which reflects the extent of clean and coal-powered energy in specific geographic locations. In the past couple of years major datacenters are offering options to select datacenter locations for computations. Whenever users are able to control this, the preferred choice should be to use infrastructure that uses cleaner energy. Further, even within a given geographic location the carbon intensity can vary depending on the time of the day\cite{Henderson2020,anthony2020carbontracker}. Using carbon tracking tools can provide insight into choosing the optimal times to schedule jobs in order to reduce the carbon emissions. Job schedulers like slurm can use this information to minimize the energy consumption.

As reported in Table~\ref{tab:distance}, the carbon emissions of using nnU-net on the two datasets with {\em AMP} is consistently lower than operating in full precision mode. Frameworks such as Pytorch\cite{paszke2019pytorch} and Tensorflow offer easy settings to operate in AMP without loss in performance\cite{micikevicius2018mixed}. 


There are several small steps users can take to reduce the carbon emissions of their model development pipeline. We summarize the discussion from above into five key points, or our THETA-guidelines for reducing carbon emissions during model development:
\begin{enumerate}
        \item {\bf T}rack-log-report for control and transparency
    \item {\bf H}yperparameter optimization frameworks instead of grid search
    \item {\bf E}nergy-efficient hardware and settings are useful
    \item {\bf T}raining location and time of day/year are important
    \item {\bf A}utomatic mixed precision training always
\end{enumerate}

In addition to these steps, open-science practices such as making pretrained models and pre-processed available can reduce the repeated computational cost and improve reproducibility in MIA. 

\vspace{-0.25cm}
\section{Conclusions}
\vspace{-0.25cm}
\label{sec:conc}
As a final note, consider MICCAI-2021 where 143/531 accepted papers had {\em segment} in their title. Many reviewers look for strong baseline comparison, and nnU-net is a thorough baseline for performing model comparison. If we assume, all these papers reported experiments on at least one 2D dataset, the total carbon emissions due to these papers would be $143\times 14.7 \approx 2102$kgCO2eq, where we use the global average carbon emissions from Fig.~\ref{fig:intensity}. However, MICCAI-2021 had an acceptance rate of about 30\%. This implies the underlying energy costs of training a baseline model on one 2D dataset is about three-fold. i.e $2102\times 3= 6306$kgCO2eq. This is the annual carbon footprint of about 27 people from a low income country, where the annual per-capita carbon footprint 236.7kgCO2eq~\cite{databank2022}! These estimates are based on very conservative estimates as most papers report more than one baseline and on multiple datasets.

Most of the suggestions in this work, including the THETA-guidelines, encourage more efficient methods or use of cleaner energy. Using more efficient methods or clean energy alone will not reduce the environmental impact of DL in MIA, as more efficient methods could result in growing demands and hence increased energy consumption. This effect, known as Jevons paradox, could nullify the gains made by improving efficiency\cite{alcott2005jevons}. Some suggestions to counteract this paradox is to use rationing or quota of resources. For instance, model selection could be constrained to the carbon budget, which was briefly addressed in \cite{selvan2021carbon}.

In conclusion, we tried to quantify the energy cost of developing medical image segmentation models on three datasets using nnU-net, which performs an exhaustive hyperparameter tuning. We presented four carbon tracking tools from literature and made a comparative assessment of their features. Based on our experiments we recommended the THETA-guidelines for reducing the carbon footprint when using DL in MIA. We hope that the community will find our suggestions on tools and practices useful; not only in reporting the energy costs but also to act on reducing them.

In line with the recommendations in this work, we report the carbon footprint of all the experiments in this work using the report generated from Carbontracker\cite{anthony2020carbontracker}:

{\em The training of models in this work is estimated to use 39.948 kWh of electricity contributing to 11.426 kg of CO2eq. This is equivalent to 94.898 km travelled by car.}\\ \\
%
{\bf Acknowledgments}
The authors would like to thank members of \hyperlink{https://ohbm-environment.org/}{OHBM SEA-SIG} community for insightful and thought-provoking discussions on environmental sustainability and MIA. 
\bibliographystyle{splncs04}
\bibliography{references}
\clearpage
\appendix
\section{Overall Carbon Emissions}

\begin{table}[h]
\caption{Quantitative comparison of the average carbon costs of different settings over the five folds, reported as the equivalent distance travelled by car, for different datasets and different regions. To demonstrate the influence of training precision we compare the full and mixed precision modes. The standard deviation is over the five-folds. Aggregate carbon emissions were reported in Table~\ref{tab:distance}}
\label{tab:distance_avg}
\centering
\scriptsize
\begin{tabular}{@{}l|cc|ccc@{}}
\toprule
\multicolumn{1}{c|}{\textbf{Dataset}}  & \multicolumn{2}{c|}{\textbf{DNK-RLT (km)} }               & \multicolumn{3}{c}{\textbf{Region (km)}}                                                                     \\ 
                                     & \multicolumn{1}{c}{{\em full}} & \multicolumn{1}{c|}{\em mixed} & \multicolumn{1}{c}{DNK\_AVG} & \multicolumn{1}{c}{EST} &  \multicolumn{1}{c}{WOR} \\ \midrule
\textbf{MONuSeg}                     &       16.6$\pm$0.1                   &  15.8$\pm$0.2                         &         13.4$\pm$0.1                &         43.9$\pm$0.2                  &      23.9$\pm$0.1                                            \\
\textbf{Heart}                       &  25.7$\pm$1.6  &   16.7$\pm$3.8                   &         21.6$\pm$1.3                  &         71.0$\pm$4.4                &      38.6$\pm$2.7                                                                       \\
\textbf{Brain}                       &         34.1$\pm$1.1                 &         32.2$\pm$3.3                  &      29.2$\pm$0.3                   &       96.0$\pm$1.1                    &               52.1$\pm$0.6                                   \\ \bottomrule
\end{tabular}
\vspace{-0.5cm}
\end{table}

Following the arguments in Sec.\ref{sec:conc}, where the total carbon cost of running baseline experiments at MICCAI-2021 was speculated per training fold, if each experiment was performed over $K$-folds then the total carbon emissions would also be easily scaled by $K$. Using a similar estimation but now scaling for the $K$ folds, the total carbon cost as equivalent distance travelled by car is given as:
\begin{equation}
    \text{ Total Distance by Car.} = N \times K \times \text{ Distance by car per training fold } / \text{Acceptance Ratio}. 
\end{equation}
Using the global average carbon cost per training fold from Table~\ref{tab:distance_avg}, N=143 from MICCAI-2021, K=5 based on default nnU-net\cite{isensee2021nnu} configuration and acceptance ratio of 1/3, for the different datasets the total carbon cost (as distance travelled) would be:
\begin{enumerate}
    \item {\bf MONuSeg}: $143\times 5\times 23.9 \times 3 = 5126.5$km travelled by car.
    \item {\bf Heart}: $143\times 5\times 38.6 \times 3 = 82797$km travelled by car.
    \item {\bf Brain}: $143\times 5\times 52.1 \times 3 = 111754.5$km travelled by car.
\end{enumerate}


\clearpage
\section{Integrating Carbontracker into nnU-net}
\label{sec:carbontracker}
\begin{listing}[h]
\vspace{-1.0cm}
\caption{Integrating {\em Carbontracker} into nnU-net for tracking the training and validation iterations. The modifications are made to the \texttt{nnunet/training/network\_training/network\_trainer.py file} commented with~\#\#\#.}
\begin{minted}[linenos,tabsize=2,breaklines,fontsize=\footnotesize]{python}
def run_training(self):
    from carbontracker.tracker import CarbonTracker
    
    ### Instantiate Carbontracker
    tracker = CarbonTracker(epochs=self.max_num_epochs,
                log_dir='logs',monitor_epochs=-1)

    ......
    
    while self.epoch < self.max_num_epochs:
    ### Start monitoring with Carbontracker 
    tracker.epoch_start()
    
    ......
    
    continue_training = self.on_epoch_end()

    epoch_end_time = time()
    if not continue_training:
        # allows for early stopping
        break

    self.epoch += 1
    self.print_to_log_file("This epoch took %f s\n" % (epoch_end_time - epoch_start_time))
    
    ### Finish monitoring an epoch with Carbontracker
    tracker.epoch_end()

    .....
    
    # now we can delete latest as it will be identical with final
    if isfile(join(self.output_folder, "model_latest.model")):
        os.remove(join(self.output_folder, "model_latest.model"))
    if isfile(join(self.output_folder, "model_latest.model.pkl")):
        os.remove(join(self.output_folder, "model_latest.model.pkl"))
        
    ### Stop tracking with Carbontracker and finish
    tracker.stop()



\end{minted}
\label{code:minimal_setup}
\end{listing}
\clearpage
\section{Learning curves and performance metric}
In the main paper, we did not discuss the performance metrics such as Dice accuracy for the different models, as our assumption was that the exhaustive hyperparameter search using nnU-net would yield the optimal configuration of the U-net. Here, we show the learning curves for the Heart dataset which was trained for 100 epochs, showing reasonable training/validation behaviour. A similar, stable behaviour was observed for the other two datasets.
\begin{figure}
    \centering
    \includegraphics[width=0.65\textwidth]{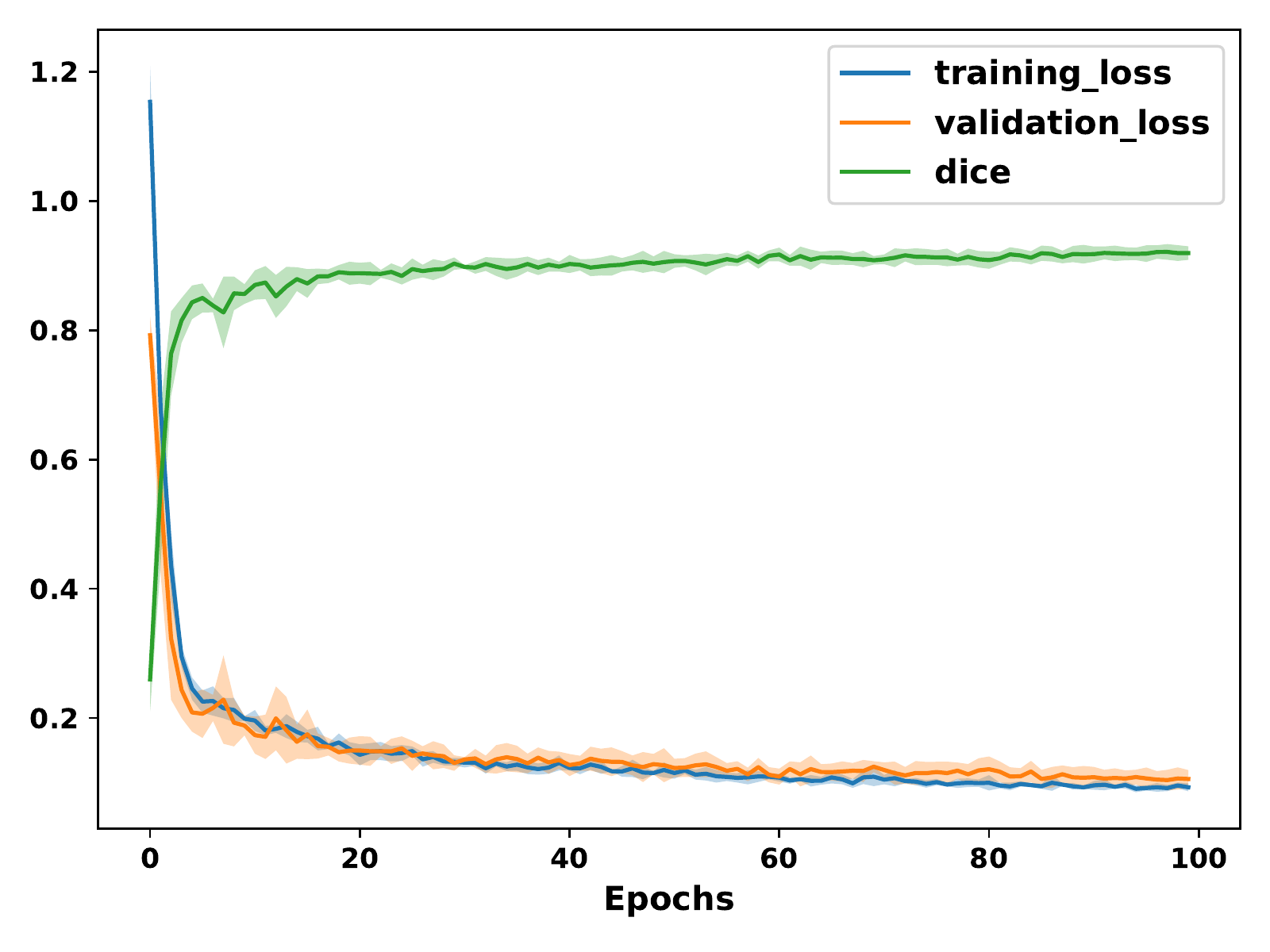}
    \caption{Training and validation losses, along with the validation dice score with standard deviation estimated over the five-folds on the Heart dataset.}
    \label{fig:learning}
\end{figure}
\vspace{-1cm}
\section{Real time carbon intensity over one year }
\vspace{-1cm}
\begin{figure}[h]
    \centering
    \includegraphics[width=0.75\textwidth]{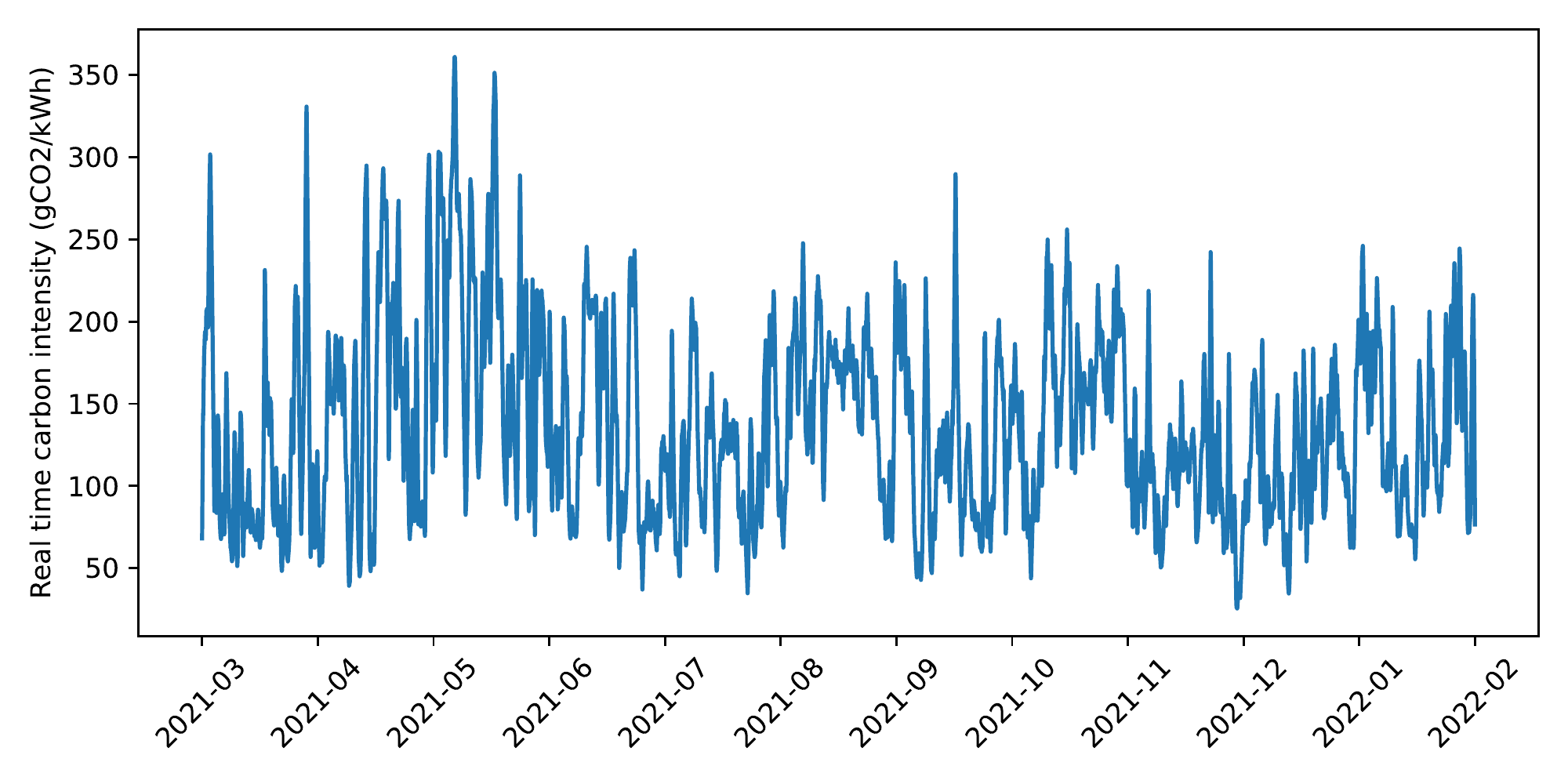}
    \caption{Real-time carbon intensity information is not publicly available for most countries, except for a handful of countries like Denmark. Here we visualize the average carbon intensity for Danish energy production and we can notice a lot of variability depending on the time of day, and time of year. Source: \url{https://www.energidataservice.dk/groups/co2-emission}}
    \label{fig:my_label}
\end{figure}


\end{document}